# Fast binarized time-reversed adapted-perturbation (*b*-TRAP) optical focusing inside scattering media

Cheng Ma[†], Fengbo Zhou[†], Yan Liu and Lihong V. Wang*

Optical Imaging Laboratory, Department of Biomedical Engineering,

Washington University in St. Louis, St. Louis, Missouri 63130-4899, USA

[†] Equal contribution *Corresponding author: lhwang@wustl.edu

**Abstract**

Light scattering inhibits high-resolution optical imaging, manipulation and therapy deep inside biological tissue by preventing focusing. To form deep foci, wavefront-shaping and time-reversal techniques that break the optical diffusion limit have been developed. For *in vivo* applications, such focusing must provide high gain, high speed, and a large number of spatial modes. However, none of the previous techniques meet these requirements simultaneously. Here, we overcome this challenge by rapidly measuring the perturbed optical field within a single camera exposure followed by adaptively time-reversing the phase-binarized perturbation. Consequently, a phase-conjugated wavefront is synthesized within a millisecond, two orders of magnitude shorter than the digitally achieved record. We demonstrated real-time focusing in dynamic scattering media, and extended laser speckle contrast imaging to new depths. The unprecedented combination of fast response, high gain, and large mode count makes this work a major stride toward *in vivo* deep tissue optical imaging, manipulation, and therapy.

**Main text**

Visible and near-infrared photons occupy a unique portion of the electromagnetic spectrum: Through non-carcinogenic molecular interaction with biological tissue and exogenous agents, they provide rich structural and physiological information[1] and noninvasive solutions for manipulation[2], control[3], and therapy[4]. However, these photons undergo severe scattering in tissue, rendering traditional control over their propagation completely ineffective beyond the optical diffusion limit — about 1 mm in biological tissue[5,6]. Such scattering shuts the door to important diagnostic and therapeutic applications at depths.

In recent years, the optical diffusion limit has been conquered by two categories of wavefront engineering techniques: wavefront shaping (WFS) and optical time reversal. In WFS[7-13], a subset of the transmission matrix[14] of the scattering medium is measured iteratively. Subsequently, the phase of the light is controlled spatially to compensate for the inhomogeneous delays due to random scattering, generating a focus through constructive interference. In optical time reversal, the optical wavefront from a real[15-17] or virtual guide star[18-20] is detected, and its phase-conjugated copy is converged back to the origin. A feasible approach to achieving time reversal is through optical phase conjugation (OPC)[21-23].

Due to the highly dynamic and scattering nature of living tissue, it is important that the focusing be sufficiently fast (ideally with a response time of ~1 ms[24,25]) and wide (i.e., having a large number of independent control elements for high focal peak-to-background ratio (PBR)[14]). Consequently, an important figure of merit is the average mode time (i.e., the average operation time per spatial mode), which should ideally approach 1 ps/mode (for a 1 ms response time with a billion modes).

However, the average mode times achieved by existing techniques are orders-of-magnitude greater than 1 ps/mode, preventing any practical applications in biological tissue. For example, WFS achieved an average mode time of 145 μs/mode (37 ms for 256 spatial modes[12]) in the fastest implementation of this technique. Ultimately, when data transfer and processing times are assumed negligible, WFS can achieve an average mode time of 18 μs/mode by employing a digital micromirror device (DMD) for wavefront contro[18,12,13]. In contrast, because OPC measures the desired wavefront at once, it is potentially faster. Analog OPC[18,21,22,25] approaches based on nonlinear optical crystals have demonstrated a very short average mode time (≈100 ps/mode, a 10 ms response time for $10^7$ modes[25]), but suffered from an energy gain (defined as the ratio of the focal light energy in the reading phase to that in the probing phase) well below unity, a drawback detrimental to wide-spread applications. In comparison, digital OPC (DOPC)[15,19,20] provides inherently large gains and reasonably fast response, with 2 μs/mode being recently realized for focusing inside scattering media[26] (100 ms response time with 50,000 spatial modes). The lowest possible average mode time (≈ 60 ns/mode) for previous DOPC techniques is dictated by the phase modulation speed (assuming a 30 ms settling time and $5\times10^5$ spatial modes), and it is still four orders of magnitude longer than the goal of 1 ps/mode. To break through this limit, the slow phase modulation must be replaced by faster binary amplitude or phase modulation[27], which could potentially reach 36 ps/mode using a DMD (assuming an 18 μs settling time[13] and $5\times10^5$ spatial modes). To date, none of the existing DOPC technologies for focusing inside scattering media have been demonstrated using binary modulation.

Here we address this challenge by demonstrating light focusing inside scattering media using binary phase modulation. The technology, named binarized time-reversed adapted-perturbation optical focusing (*b*-TRAP), has achieved the shortest average mode time to date (≈ 300 ns/mode) among all methods developed for focusing inside scattering media with high energy gain. Compared with the binary-amplitude WFS technology[13] that finds and utilizes the same binary mask, our technology uses an adaptive mask, enhances the speed by $10^5$ times, and doubles the focusing quality at no additional cost. More strikingly, we show that by using *b*-TRAP, the formation time of the binary mask is determined only by the laser pulse interval, which is highly tunable and can be less than 1 μs. These features not only facilitate fast light focusing, but also enable us to acquire detailed information about tissue dynamics (such as blood flow speed). The demonstrated capabilities pave the way toward *in vivo* deep tissue biophotonics.

**Principles**

Time-reversed adapted-perturbation (TRAP) optical focusing[26,28] employs intrinsic tissue dynamics as guide stars for completely noninvasive and non-contact light focusing inside scattering media. In short, if the scattered fields at two instants are recorded in the presence of internal dynamic perturbations (e.g., movements and absorption or refractive index changes), subtracting the two fields generates a differential field, whose conjugate copy enables focusing at the perturbed sites by cancelling the scattering contribution originating from the static portion of the medium.

The original TRAP focusing[28] relied on phase-shifting holography[29] to record the complex amplitude of the scattered electromagnetic fields, which took multiple intensity measurements to accomplish and was impractically slow for many applications. In comparison, *b*-TRAP measures each field indirectly within the time duration of a single laser pulse. As shown in Fig. 1**a**, at instant $t_1$, a short laser pulse probes the scattering medium in which the permittivity of the target (i.e., the guide star) is $\varepsilon(t_1)$. The exterior scattered light field, termed the "sample beam", whose complex amplitude is expressed as $E_S(t_1)$, is combined with a planar reference beam $E_R$ on an amplitude-only spatial light modulator (SLM). The SLM surface is imaged onto a scientific complementary metal oxide semiconductor camera (sCMOS) to generate a time-averaged intensity pattern

$$I(t_1) = |E_S(t_1) + E_R|^2 = |E_S(t_1)|^2 + |E_R|^2 + E_S^*(t_1) E_R + E_S(t_1) E_R^*. \quad (1)$$

As shown in Fig. 1**b**, at a later instant $t_2$, the above process is repeated, at which time the target permittivity has changed to $\varepsilon(t_2)$ either from naturally occurring event, such as movement, or from externally induced perturbation. The result is a different intensity pattern:

$$I(t_2) = |E_S(t_2) + E_R|^2 = |E_S(t_2)|^2 + |E_R|^2 + E_S^*(t_2) E_R + E_S(t_2) E_R^*. \quad (2)$$

Subtracting the two patterns (equations (1) & (2)) generates multiple terms:

$$\Delta I = I(t_2) - I(t_1) = \Delta|E_S|^2 + \Delta E_S^* E_R + \Delta E_S E_R^*, \qquad (3)$$

where $\Delta|E_S|^2 = |E_S(t_2)|^2 - |E_S(t_1)|^2$, and $\Delta E_S = E_S(t_2) - E_S(t_1)$. The second term on the right-hand side of equation (3) encodes the TRAP field $\Delta E_S^*$, which can be reconstructed upon reading an amplitude hologram $\Delta I$ with a conjugate reference (reading) beam $E_R^*$ (Fig. 1c). It should be noted that $b$-TRAP focusing dramatically simplifies the complicated task of complex amplitude measurement, and allows indirect, single laser-shot electric field recording. The advantages of the new technology will be discussed in detail in the next section.

The above focusing procedure can be implemented using the schematic shown in Fig. 1d. The system employs a Mach-Zehnder interferometric structure where the scattering medium resides in one arm and the planar reference/reading beam is generated in the other. Intensity patterns $I(t_1)$ and $I(t_2)$ are recorded via an sCMOS camera. The SLM, along with the camera, forms a DOPC system[23]. Adjusting the driving frequencies ($f_1$ and $f_2$) of the two acousto-optic modulators (AOMs) allows the system to work under different operation modes, as detailed in the next section. For a more detailed description of the system setup, see Methods.

**Operation modes**

TRAP relies on embedded novelties[30] to guide light focusing, meaning that the target's permittivity must be time-variant. Accordingly, the light field scattered by the target continuously decorrelates over time[31]. In real practice, the background medium (e.g., biological tissue) also generates time-variant scattering, resulting in a continuously decorrelating background field. The focusing quality sensitively hinges on the background field decorrelation during the entire time-reversal process[25,32]. Hence, TRAP focusing is valid only if the field from the target decorrelates faster than the field from the surrounding medium. The time interval between the field measurements must be properly chosen to allow for detectable target decorrelation yet a highly correlated background. The above condition is illustrated in Fig. 2**a-c**. The curves drawn on the upper rows depict distinct correlation decays derived from the target's and the background's scattering. During the process, the difference between the two field measurements exposes the target decorrelation against a relatively stable background. For example, the correlation time of flowing blood is on the order of milliseconds, while that of the surrounding tissue varies, but is at least 10 times slower[22,25,33] with proper stabilization.

Note that the duration of each field measurement must be shorter than the target decorrelation time to avoid significant reduction of focal PBR. As shown in Fig. 2**a**, the original TRAP focusing scheme relied on digital phase shifting holography, which required at least four laser shots synchronized with four camera exposures to complete a single field measurement. The speed of the method was primarily restricted by the camera frame-rate. For example, if the camera ran at 50 Hz, the minimum duration of four exposures was ~ 60 ms. Consequently, the target needed to be stable within 60 ms, which is impractically long for many applications. Moreover, the second field measurement had to be significantly delayed (e.g., to after 1 s) due to the slow target decorrelation, making the method applicable only to a nearly stable background.

To remove such limitations, the field measurement time must be shortened. As shown in Fig. 2**b**, *b*-TRAP records each of the two fields with a single laser shot, which is accomplished on a time scale defined by the laser pulse width. By using a laser with 6 ns pulse width, the original constraint on the target correlation time is relaxed by 7 orders of magnitude (from 60 ms to 6 ns[28]). Such a fast field measurement makes *b*-TRAP applicable to almost any biological targets. The phase differences between the reference beam and the static portion of the sample beam are kept constant for the two measurements, and thus digitally subtracting the two intensity patterns directly generates the TRAP field (equation (3); see Methods for details). The response speed of the scheme is limited only by the camera frame rate and SLM refresh time. Because it requires two camera exposures, the scheme is named double-exposure *b*-TRAP focusing.

*b*-TRAP focusing has been made even more powerful by the single camera-exposure scheme described below (referred to as single-exposure *b*-TRAP hereafter), which can reduce the hologram formation time by two orders of magnitude in comparison to the state of the art[26]. As shown in Fig. 2**c**, in this mode, one camera exposure records two consecutive laser shots, with a π shift introduced between the sample and reference beams in the second shot (see Methods for details), resulting in a total intensity of

$$I = \Sigma |E_S|^2 + 2|E_R|^2 + \Delta E_S^* E_R + \Delta E_S E_R^*,$$

where $\Sigma |E_S|^2 = |E_S(t_1)|^2 + |E_S(t_2)|^2$, and the TRAP field is encoded in the third term. In this focusing mode, the time interval between the two field measurements is determined by the laser repetition rate, making the procedure independent of the camera frame rate. Thus, a hologram can be formed within less than 1 μs for operation at a >1 MHz laser repetition rate (e.g., by using a mode-locked laser). The ultimate speed is determined only by the image transfer and SLM refresh rates.

The advantage of *b*-TRAP focusing is demonstrated by the simulation results shown in Fig. 2**d-f**. In these simulations, a point target moves (downward) inside a dynamic scattering medium. The original TRAP scheme (Fig. 2**a**) fails to focus (Fig. 2**d**) due to its low speed. In vivid contrast, *b*-TRAP focusing successfully captures the target movement and focuses into the time-variant scattering media with double exposures (Fig. 2**e**) and a single exposure (Fig. 2**f**). However, single-exposure *b*-TRAP shows the strongest background suppression (Fig. 2**f**), thanks to its higher speed and greater tolerance to medium decorrelation. See Supplementary Method 1 for details.

**Hologram binarization**

A *b*-TRAP hologram can be treated as a superposition of random amplitude gratings, whose diffraction efficiencies are highly dependent on their peak-to-peak variation of amplitude reflectance relative to the background. The average relative reflectance variation is small in our case, resulting in weak focusing intensity. However, the average reflectance variation can be enhanced as follows—detailed in Supplementary Method 2. First, the mean is filtered out. Second, the resultant hologram is multiplied by a factor $K$. Third, a constant background of 0.5 is added to the hologram. Finally, the pixel values are truncated at zero and unity. At sufficiently large values of $K$, most pixel values are truncated, giving birth to a binary hologram.

The amplification process with increasing $K$ is shown in Fig. 3**a**. With $K = 16$, the hologram becomes nearly binary, with most elements being either 0 or 1, as confirmed by the gray value probability density plot shown in Fig. 3**b**. The focus and background intensities at various $K$ values are shown in Fig. 3**c** (i) & (ii), and both simulation and experimental results confirm that the diffraction efficiency is effectively improved, see Supplementary Method 2 for experimental and simulation details. On the contrary, the nonlinear truncation process is accompanied by reduced wavefront reconstruction fidelities, and accordingly, the focal PBR is expected to drop[13,20]. However, both experiment and simulation indicate that the PBR actually undergoes a fast surge before it slightly drops (see Fig. 3**c** (iii)). This phenomenon can be explained by the leakage of the zero-order diffraction pattern of the time-reversed light, explained in more detail in Supplementary Method 2. The results shown in Fig. 3**c** demonstrate that a binary amplitude hologram leads to near-optimal focusing in both peak intensity and PBR (see Discussion for quantitative comparisons). We used a zero-order block to double the focal PBR by converting binary amplitude to binary phase modulation[27], see Methods for details. The practical significance inherent in this observation will be analyzed in the Discussion. Subsequently, we used binary amplitude holograms throughout the rest of the experiments.

**Moving object tracking**

Real-time tracking of moving objects has extensive applications[34,35], and TRAP focusing is the only known optical method that tracks moving objects inside scattering media. A robust tracking system must meet two criteria. First, the time it spends on field measurement must be sufficiently short. Since there is no external control over the objects' moving speeds, it is essential that the field measurement be accomplished sufficiently fast to accommodate potentially fast field decorrelations. Second, the entire focusing procedure, including measurement, calculation, and display of the hologram, must be finished within a short time span so as to minimize the spatial lag of the focus, and to robustly adapt to any potential host medium decorrelation.

In a demonstration of the original TRAP focusing, the target moved slowly, with an estimated decorrelation time of ~6 s. With the new *b*-TRAP focusing system, we shortened the field measurement time from 60 ms down to 6 ns, and reduced the repetition period from ~2 s to 140 ms. These speed improvements allowed us to optically track fast moving objects inside scattering media. A simplified experimental setting is depicted in Fig. 4**a**, where a black human hair target (~50 μm in diameter), vertically positioned between two optical diffusers (DG10-600, Thorlabs, USA), was mounted onto a motorized stage. The target was moved back and forth in the *x* direction at a speed of 0.78 mm/s (with an estimated decorrelation time of <60 ms), while *b*-TRAP focusing was performed at a repetition rate of 7 Hz. A detailed explanation of the timing of the focusing procedure, including key steps and their execution times, are provided in Supplementary Method 3. The focal light intensity distribution was sampled using a beamsplitter onto the detection plane of a CMOS camera (Firefly MV, Point Grey, Canada) to monitor the focusing process in real time. The two-dimensional (2D) light intensity distribution was then averaged along *y*, condensed into a 1D intensity map, and stacked in time to form an intensity distribution time-trace, as shown in Fig. 4**b**. The movement of the target was clearly revealed by the tracking focus. Most of the focusing patterns comprised two foci, typical of TRAP focusing patterns induced by a moving object[28]. As shown in the inset of Fig. 4**b**, the twin foci were contributed by the old and the new target positions.

**Tissue-mimicking phantom experiment**

One unique feature of TRAP focusing is its ability to concentrate light onto endogenous contrast agents, such as moving red blood cells (RBCs). As RBCs are exclusively confined within blood vessels, TRAP focusing selectively deposits time-reversed photons within the vasculature. This feature is potentially useful in applications such as photoacoustic tomography of blood vessels[36] and treatment of port-wine stains[37]. It is important to note that live tissue is associated with two distinct correlation times: A fast decorrelation due to blood flow, and a slow one from the relatively stationary background[25,32]. The TRAP focusing procedure must be able to capture the fast changing information, and to adapt to the slow background evolution.

We demonstrated the above capability in the tissue-mimicking phantom experiment shown in Fig. 5**a**. The phantom was composed of two ground-glass diffusers (DG10-600, Thorlabs, USA) as scattering media with a tube containing flowing blood (300 μm inner diameter, silicone) placed vertically in between. The mimic blood vessel was completely invisible outside of the ground glass. To mimic tissue decorrelation with controlled correlation time, the scattering layers were mounted onto a motorized stage with adjustable moving speed. The field correlation time[25] of the scattered light was measured as a function of the media's movement speed, and subsequently fitted according to

$$t_C[\text{s}] = 1.2[\mu\text{m}] \times \{v[\mu\text{m/s}]\}^{-1}, \tag{4}$$

as shown in Fig. 5c. From equation (4), we were able to pre-set the tissue correlation time by controlling the media's movement speed. During the focusing demonstration, diluted bovine blood (1 (blood):8 (PBS solution) by volume) was pumped through the silicone tube at a constant flow rate. The moving RBCs perturbed the scattered field, forming targets for time-reversed focusing of the entire vessel. While the focusing procedure was performed at a repetition rate of 7 Hz, the scattering media were translated at various speeds to set the background correlation time to $+\infty$, 0.81 s, 0.41 s, and 0.28 s; the corresponding 2D focal light intensity distributions and 1D cross-sectional intensity profile along $x$ are plotted in Fig. 5 **b** and **d** in subplots (i)-(iv), respectively. Despite an intensity drop, the focus was well preserved, even when the background medium decorrelated within as short as 0.28 s. Such demonstrated capabilities take a quantum leap from an earlier demonstration[28], where the slow response forced the blood flow to be temporarily stopped to facilitate field measurements, and required absolutely stationary background media.

**Deep flow measurement**

As shown in the preceding section, blood flow perturbs the scattered light field, which enables time-reversed focusing: The field decorrelation associated with the RBC movement is captured externally and converted to a differential field, the conjugate copy of which ultimately leads to the time-reversed focusing. Two consecutive laser pulses "see" two sets of spatially shifted RBC positions, as illustrated in Fig. 6**a**. The displacement $d = v \times \Delta t$, where $v$ is the flow speed and $\Delta t$ is the pulse interval. A larger displacement results in greater decorrelation (Fig. 6**d**), creating a stronger differential field (Fig. 6**b**) and a brighter focus (Fig. 6**c**). The focusing power ($P_{TRAP}$) is a function of RBC displacement:

$$P_{TRAP} = P_{max} \left\{ 1 - \left\{ 2\cos^{-1}\left(\frac{d}{D}\right) - \sin\left[2\cos^{-1}\left(\frac{d}{D}\right)\right] \right\} \Big/ \pi \right\}, \qquad (5)$$

where $D$ is the average diameter of the moving particle and $P_{max}$ is the maximum power, see Supplementary Method 4 for detailed analysis.

At fixed $\Delta t$, $d$ was adjusted by varying $v$, giving rise to dramatically different focal PBR. Experimental results are compared with theoretical ones for $\Delta t = 2$ ms (500 Hz laser repetition rate) and $\Delta t = 1.25$ ms (800 Hz laser repetition rate) in Fig. 6**e**. The theoretical curves were plotted based on equation (5), with $P_{max}$ and $D$ fitted to the experimental data. The effective RBC diameters were found to be 3.4 μm and 3.2 μm for $\Delta t = 2$ ms and 1.25 ms, respectively; the difference is attributed to measurement noise (See Methods for details). For flow measurement, the laser repetition rate can be swept to scan $d$, and $v$ is then readily obtained by fitting. The focal PBR and location can be acquired photoacoustically[38].

In the experiment, the interval between the two field measurements must be precisely adjustable and can be squeezed below the target decorrelation time. Among all technologies developed to date, only single-exposure *b*-TRAP focusing has this unique capability. It is important to note that in such working mode, the time to generate a hologram is entirely dependent on the laser pulse interval (as opposed to the camera frame-rate), which is short and highly adjustable. Consequently, the finely and broadly tunable $\Delta t$ makes the method extremely appealing due to its large dynamic range: It can measure both slow[39] and fast[40] flow, and extends laser speckle contrast imaging of blood flow[31] into the diffusive regime.

**Discussion**

We discovered that the *b*-TRAP mask is equivalent to the hologram generated by the binary amplitude WFS technique[13], but we doubled the PBR by employing a zero-order block for binary phase modulation[27] (see Supplementary Discussion 3 for details). Using a DMD, the binary amplitude WFS technique took several minutes to find the mask with 3228 degrees of freedom (DOF; i.e., controlled number of spatial modes); the limited speed was attributed to the iterative algorithm and data transfer between a personal computer and the DMD. In comparison, *b*-TRAP focuses in 143 ms with a DOF of $5\times10^5$, equivalently $10^5$ faster than the WFS approach on a per mode basis.

In OPC, two types of systems were developed. The analog systems perform optical phase conjugation using nonlinear crystals. The response of such systems can be fast, ranging from milliseconds[25] to microseconds[41]. The DOF of the analog systems are large[25], and can exceed $10^7$. Nevertheless, the gain of the systems is low. Despite a large power gain recently demonstrated using a pulsed hologram readout[42], the achievable energy gain is still well below unity. Without large energy gains, many applications are still infeasible.

In digital systems, optical phase conjugation is accomplished by digital spatial light modulators, which can provide high gains[20]. However, the low camera frame rates significantly limit the field measurement times and the overall speeds. A single-shot off-axis holographic scheme has been employed to reduce the field measurement time and accelerate focusing[32]. However, the speed improvement compromises spatial resolution and reduces the total number of DOF significantly (see Supplementary Discussion 1 for details). In contrast, *b*-TRAP focusing achieves a similar field measurement time (enabled by single-shot holographic recording) while preserving the number of DOF; thus, the technology has the shortest average mode time among all digital systems developed to date (We have demonstrated an average mode time of ~0.3 μs/mode, using a focusing repetition rate of 7 Hz and $5 \times 10^5$ spatial modes). Moreover, for double-exposure *b*-TRAP focusing, the computational load is significantly reduced compared to other digital approaches. It took ~30 ms on a 3.6 GHz quad-core i7 CPU (Intel, USA) for Matlab to generate a binary mask with 1920 × 1080 resolution. The single-exposure scheme further improves the focusing speed by automatically forming the TRAP hologram within one camera exposure, thereby decoupling the speed from the camera frame-rate and hologram computation time, leaving the ultimate speed defined by the rates of data transfer and SLM actuation. In deep flow measurements, we have achieved a hologram formation time of 1.25 ms, which is nearly two orders of magnitude shorter than that of the state-of-the-art technique[26]. By using a pulsed laser with >1 MHz repetition rate, the hologram formation time can be further reduced to below 1 μs, sufficiently fast for most biological applications. Moreover, this unique capability allows us to not only detect, but also quantify target decorrelation during time-reversal (which is the key to deep flow measurement).

The demonstrated binary modulation makes *b*-TRAP implementable on DMDs with minimal system modification, which can potentially further improve light focusing speed by 3 orders of magnitude, making *in vivo* applications possible. Currently, besides low speed, the number of DOF supported by the liquid-crystal-based SLM is insufficient for most real applications, since the resultant low focal PBR in deep tissue is unsuitable for applications such as imaging, therapy or control. On the other hand, the DMD has recently emerged as an attractive solution for optical time reversal thanks to its high speed[8,12,13,27,43-47], and its relatively low price makes multiplexing affordable, which can potentially increase the total number of DOF dramatically. The method invented herein can be integrated with other time-reversal based approaches, such as ultrasonically encoded focusing (TRUE)[18-20], to improve their speeds and efficiencies.

Single-exposure *b*-TRAP focusing has a smaller diffraction efficiency than the double-exposure approach because the ratio of the TRAP component in the hologram is lower. As dictated by the nature of amplitude modulation, *b*-TRAP schemes have weaker diffraction compared to those focusing methods implementing phase modulations; the focusing efficiency is about 10 times lower (see Supplementary Discussion 2 for details). It is important to note that lower diffraction efficiency does not necessarily lead to lower PBR (which determines the focusing quality). The light reflected as the zero-order component can be effectively blocked by a spatial filter, leading to binary phase modulation with a focal PBR ~40% of that attainable using a phase-only approach (see Supplementary Discussion 3 for details). To compensate for the lower reflectivity, stronger input light can be used as long as damage to the SLM is avoided. Multiple DMDs can work in parallel to further enhance the energy gain.

In sum, optical time reversal techniques promise to revolutionize biomedical optics by dramatically extending the depth of optical focusing inside tissue. However, several key factors, including gain, speed, and focal PBR, must be simultaneously improved to implement these techniques *in vivo*. *b*-TRAP focusing has shown great potential in fulfilling all these criteria and in bridging the gap between laboratory explorations and real-world applications of deep tissue biophotonics. The technology has anticipated applications in deep-tissue molecular imaging[48], flow measurement and cytometry[26], photodynamic therapy[49], optogenetics[50], micro-surgery[4], and more.

**Methods**

**Setup.** The system is schematically shown in Supplementary Fig. S1. A Q-switched 532 nm laser (Customized, Elforlight Inc., UK) produced 6 ns, 0.6 mJ pulses with a repetition rate tunable from 500 Hz to 1000 Hz. The coherence length of the laser was measured to be > 0.5 m, with sporadic fringe instability due to mode hopping. The pulse train entered a Mach-Zehnder interferometer through a polarizing beamsplitter (PBS) with a splitting ratio adjustable via a half-wave plate (HWP). Two acousto-optic modulators (AOMs, AFM-502-A1, IntraAction, USA) were cascaded in the sample arm (which is slightly different from the schematic shown in Fig. 1, but the function is unchanged) to provide tunable optical frequency shifts. The frequency-shifted light illuminated a scattering sample, with the scattered light collected on the other side in a transmission configuration. The $f$ = 35 mm plano-convex lens used to collect the scattered light imaged the back side of the scattering sample to the surface of an amplitude-only SLM (HED 6001, Holoeye, Germany) having 8 bit graylevel and 1920 × 1080 resolution. An optical polarizer was used in conjunction with the SLM for best amplitude modulation contrast. The SLM surface was conjugated to the image plane of an sCMOS camera (pco.edge, PCO AG, Germany) by a camera lens (Nikon, Japan) with 1:1 digital pixel matching. The light in the other arm went through another HWP to rotate its polarization by 90º. The light was further split into a reference and a reading beam by a 50:50 beam splitting plate. The reference beam was attenuated ten times by a neutral density filter and properly delayed to minimize mode-hopping-induced fringe instability. The reference and reading beams were carefully aligned to be coaxial after being combined by a 90 (reflection):10 (transmission) BS. They were collimated and expanded by an afocal system to 1 inch in diameter before being merged with the sample beam via a 50:50 BS. Switching the sample and reading beams "on" and "off" was controlled by two optical shutters (Uniblitz LS3, Vincent Associates, USA), which, together with the two AOMs and the sCMOS camera, shared the same timebase with the laser Q-switch. During hologram measurement, shutter 1 was open while shutter 2 was closed, resulting in an interference pattern formed by the sample and reference beams. After two intensity patterns were captured by the sCMOS camera and subsequently processed, a TRAP hologram was calculated, binarized, and displayed on the SLM. In the time-reversed focusing mode, shutter 1 was closed and shutter 2 was open, allowing a strong reading beam to reconstruct the TRAP light. To improve the focal PBR, the strong retro-reflected zero-order component was effectively blocked by a black dot (printed on a transparency) placed at the focus of the $f = 35$ mm light-collecting lens. The zero-order block converted binary amplitude to binary phase modulation, see Supplementary Discussion 3 for details. Inside the scattering sample, the light intensity distribution on the focal plane (target plane) was mirrored to the image plane of a CMOS camera (FireflyMV, Point Gray, Canada) by a BS if real-time monitoring of the focus was needed.

**Double- and single-exposure synchronization schemes.** In double-exposure $b$-TRAP focusing, the modulation frequencies of AOM1 and AOM2 are set to $f_1 = 50$ MHz (50 MHz is the central frequency of the device) and $f_2 = -f_1 + \Delta f$. The frequency difference is $\Delta f = N f_C$, where $f_C$ is the frame-rate of the camera and $N$ is a non-negative integer. The laser repetition rate is $f_L = M f_C$, where $M$ is a positive integer. In the single-exposure scheme, the sign of the beat between the sample and reference beams flips from shot to shot, requiring $\Delta f = (N+1/2)f_L$, where $N$ is a non-negative integer. Laser shots are synchronized with camera exposures, with $f_L = Mf_C$ ($M > 2$) and $2/f_L < t_{exp} < 3/f_L$, where $t_{exp}$ is the camera exposure time.

**Deep flow measurement procedure.** Blood flow speed was controlled by a motorized syringe pump during the measurement. We used a hybrid scheme to generate the digital hologram for the best focal PBR. Two laser repetition rates were employed during the test: $f_L$ = 500 Hz and $f_L$ = 800 Hz. For $f_L$ = 500 Hz, $f_C$ = 12.5 Hz, $\Delta f$ = 250 Hz, and $t_{exp}$ = 3 ms. For $f_L$ = 800 Hz, $f_C$ = 12.5 Hz, $\Delta f$ = 400 Hz, and $t_{exp}$ = 2 ms. At a fixed flow speed, 51 single-exposure b-TRAP holograms were recorded, and the hybrid holograms were calculated using $H_{H,i} = H_{SE,i+1} - H_{SE,i}$ ($i$ = 1,…,50), where $H_{H,i}$ represents the $i$-th hybrid hologram, and $H_{SE,i}$ denotes the $i$-th single-exposure hologram. The focal PBR was then characterized using a CMOS camera (Firefly MV, Point Gray, Canada) by averaging over the 50 corresponding focal light patterns. Repeating the above procedure at various flow speeds produced the experimental results shown in Fig. **6e**. To finalize the characterization process, the results were then fitted using a nonlinear least-squares model to equation (5), with $P_{max}$ and $D$ as unknown parameters.

**Acknowledgements** We thank Song Hu for discussions about the flow measurement experiment, and James Ballard for editing the manuscript. This work was sponsored by National Institutes of Health grants DP1 EB016986 (NIH Director's Pioneer Award) and R01 CA186567 (NIH Director's Transformative Research Award).

**Author contributions** C.M., F.Z. and L.V.W. initiated the project. F.Z. and C.M. built the *b*-TRAP focusing system. C.M., F.Z. and Y.L. designed and conducted the experiments. C.M. wrote the codes for the experiments and simulation, and processed the experimental data. L.V.W. provided overall supervision. C.M. wrote the manuscript, and all authors involved in editing the manuscript.

**Author information** Reprints and permissions information is available at www.nature.com/reprints. The authors declare no competing financial interests. Correspondence and requests for material should be addressed to L.V.W. (lhwang@wustl.edu).

**Figure captions**

**Figure 1.** Principle and schematic of *b*-TRAP focusing. **a-c**, Principle of *b*-TRAP focusing (details see text). **d**, Schematic of the focusing method. Green (solid) arrows represent light path in the probing process; blue (slim) arrows show light path in the time reversal process. AOM, acousto-optic modulator; BS, beam-splitter; CL, camera lens; L, lens; M, mirror; OS, optical shutter; sCMOS, scientific complementary metal oxide semiconductor camera; SF, spatial filter; SLM, spatial light modulator.

**Figure 2.** Influence of field sampling scheme on the focusing quality under both target and medium decorrelations. **a-c**, Time-dependent correlation coefficients of the scattered electromagnetic fields due to the target and the background (upper rows), and the field sampling schemes (lower rows). **a**, Full-field (amplitude and phase) sampling using a phase-shifting scheme. **b**, Amplitude-only sampling with double camera exposures and digital subtraction. **c**, Amplitude-only sampling with single camera exposure and automatic analog subtraction. **d-f**, Simulated focal light intensity distributions corresponding to the sampling schemes in **a-c**, normalized by the peak intensity in **f** (for details see text).

**Figure 3.** Effect of binarization of the TRAP hologram. **a**, Hologram binarization process, showing the same portion of the hologram (pseudo 3D view) at two amplification factors ($K = 1$ and 16). The holograms are truncated between gray values of 0 and 1. **b**, Probability density function of the gray value in holograms with varied $K$. **c**, Focusing quality as a function of $K$, showing the evolution of the peak intensity (i),

background intensity (ii), and focal peak-to-background ratio (PBR) (iii) on the focal plane. Dots are experimental data, and curves are simulation results.

**Figure 4.** Tracking a moving target inside scattering media. **a**, Experimental scheme. The hair target between two scattering media moves back and forth on a motorized stage. **b**, Measured time trace of the focal light intensity distribution. Each row depicts the light intensity distribution (integrated along *y*) at a fixed time point. Each column shows the temporal evolution of the intensity at a fixed position. Process repetition rate, 7 Hz. Insets: Intensity profile along the dashed line.

**Figure 5**. Tissue-mimicking phantom experiment. **a**, Experimental arrangement. RBC, red blood cell; BS, beam-splitter. **b**, Focal light intensity distribution, each averaged over 100 speckle realizations, measured at different movement speeds of the media. **c**, Medium's correlation time as a function of its movement speed. Triangles are measured data, solid line is the fitted curve. **d**, Evolution of the focal intensity profile with increasing media decorrelation rate. Indices i through iv in **b-d** label different medium's correlation times as shown in **c**. Scale bar, 500 μm.

**Figure 6.** Deep flow measurement with single-exposure *b*-TRAP focusing. **a**, Simplified schematic. *d*, displacement; *D*, size of the moving particle. **b**, Intensity maps of the exterior differential field (upper row, simulation) and interior focal field (lower row, experimental) at various target displacements. **c**, Correlation of the scattered fields at different target displacements. Stars, simulation; dashed curve, theory. **d**, Flow measurement results showing focal peak-to-background ratio (PBR) as a function of flow

speed, obtained at 500 Hz and 800 Hz laser repetition rates, respectively. Discrete points are experimental data, curves are theoretical fittings. Scale bar, 500 μm.

Figure 1

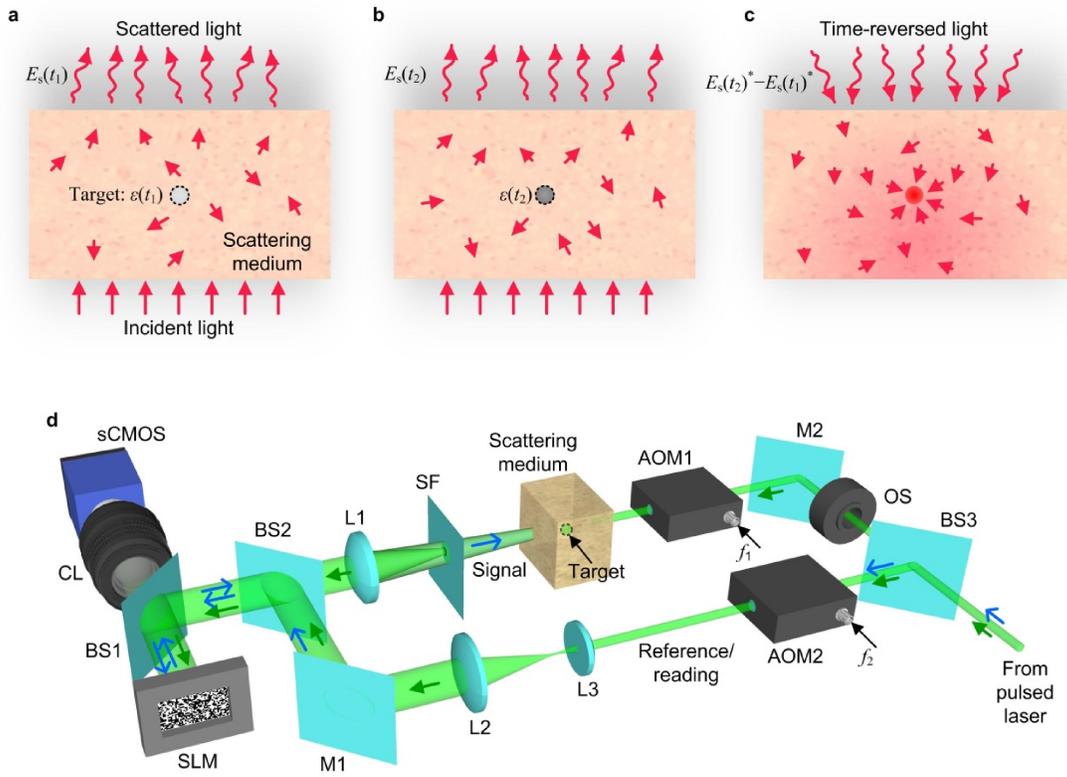

Figure 2

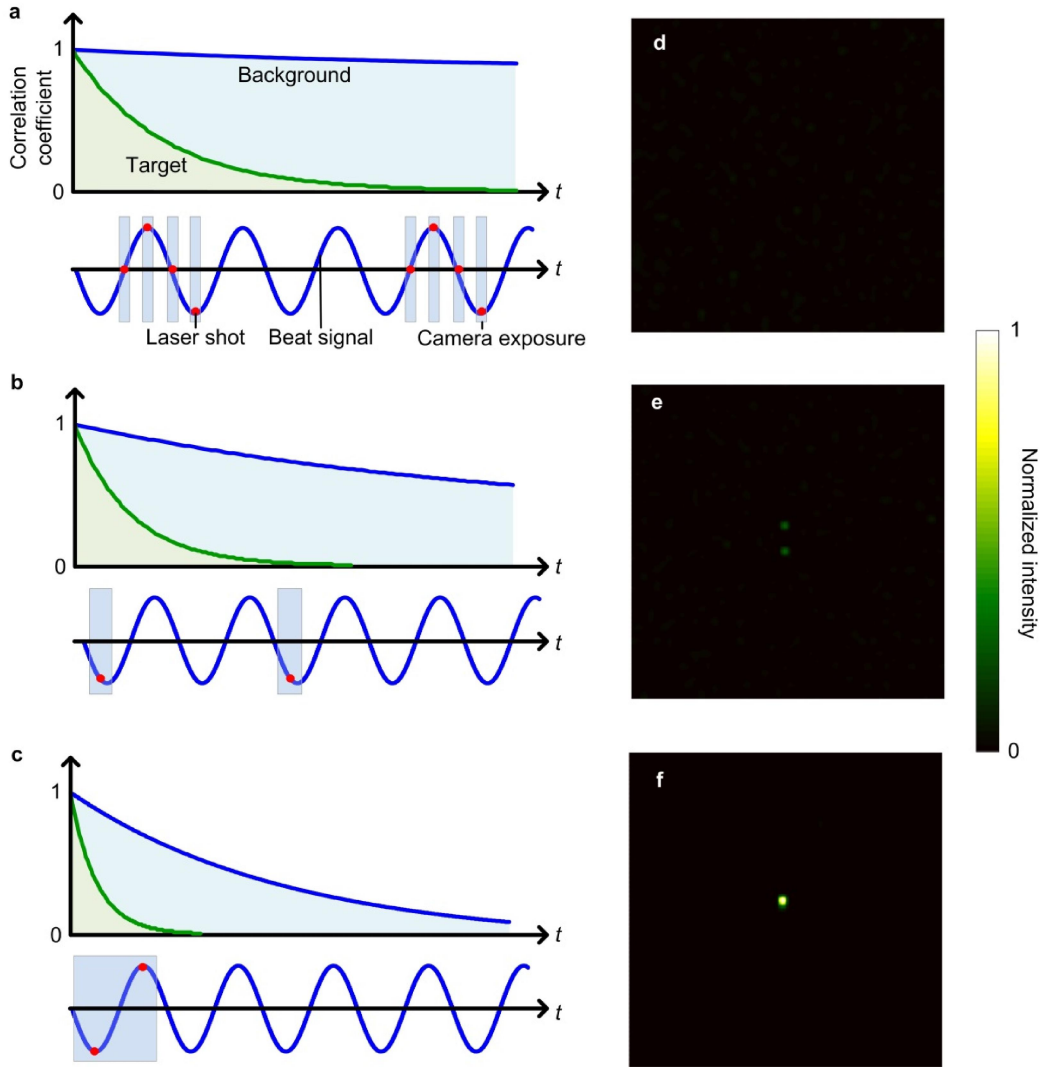

Figure 3

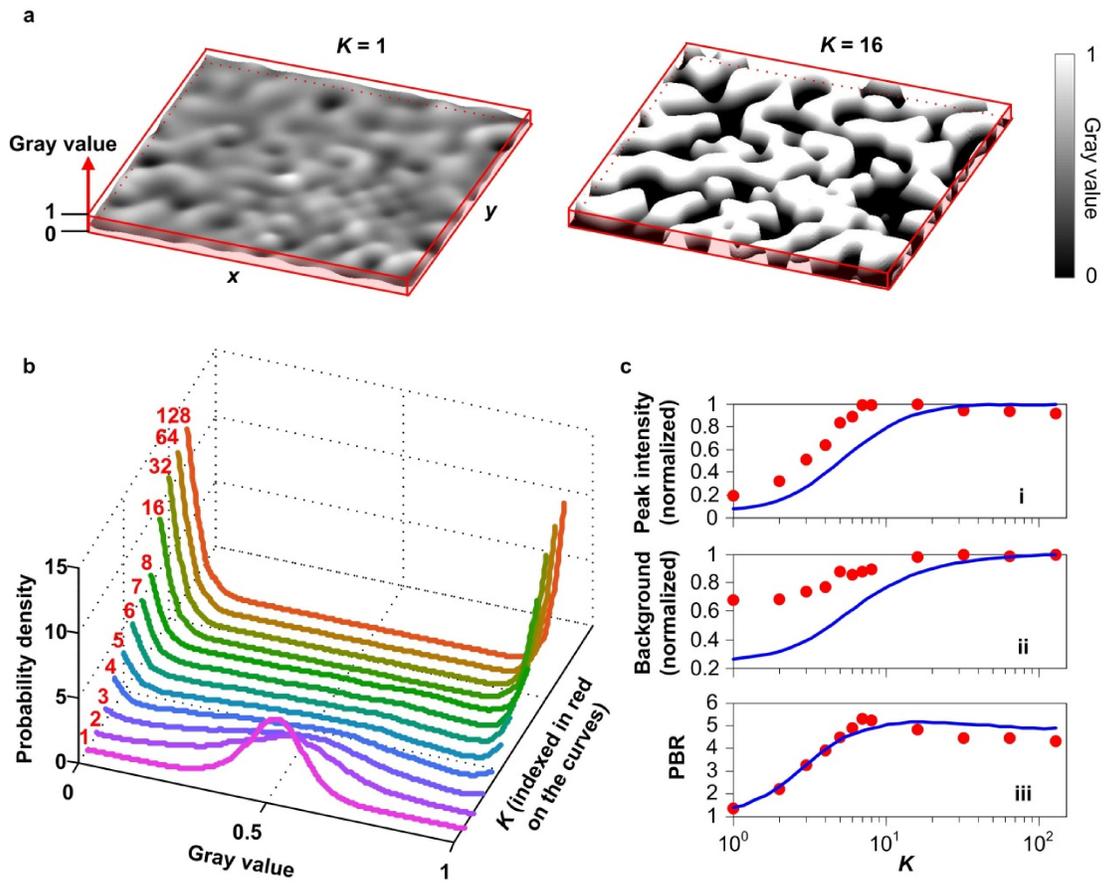

Figure 4

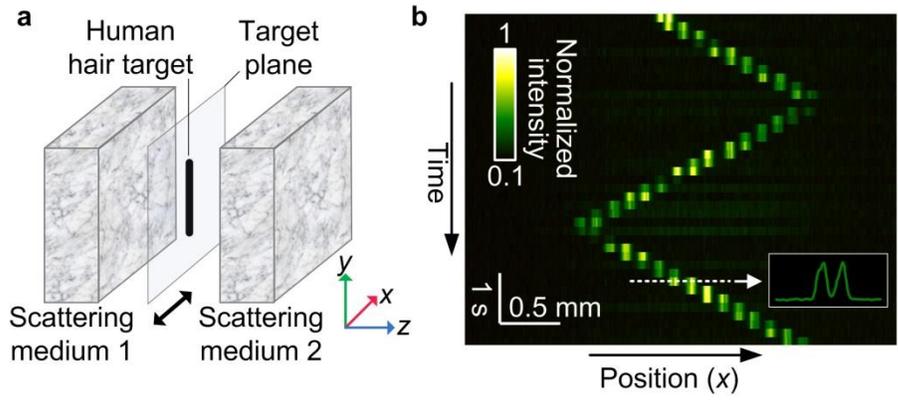

Figure 5

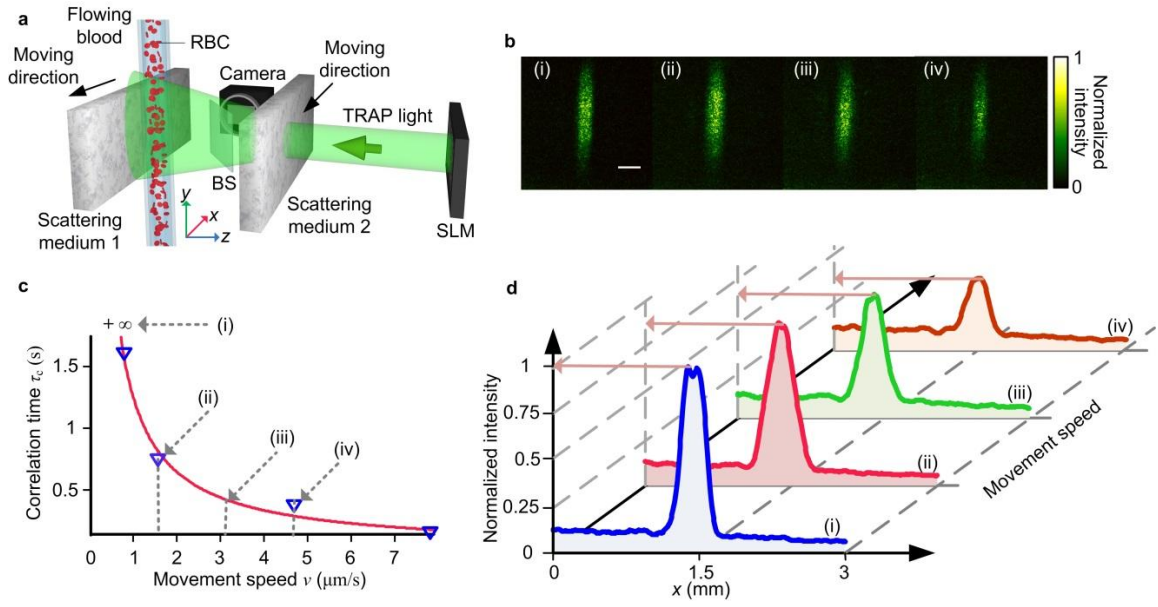

Figure 6

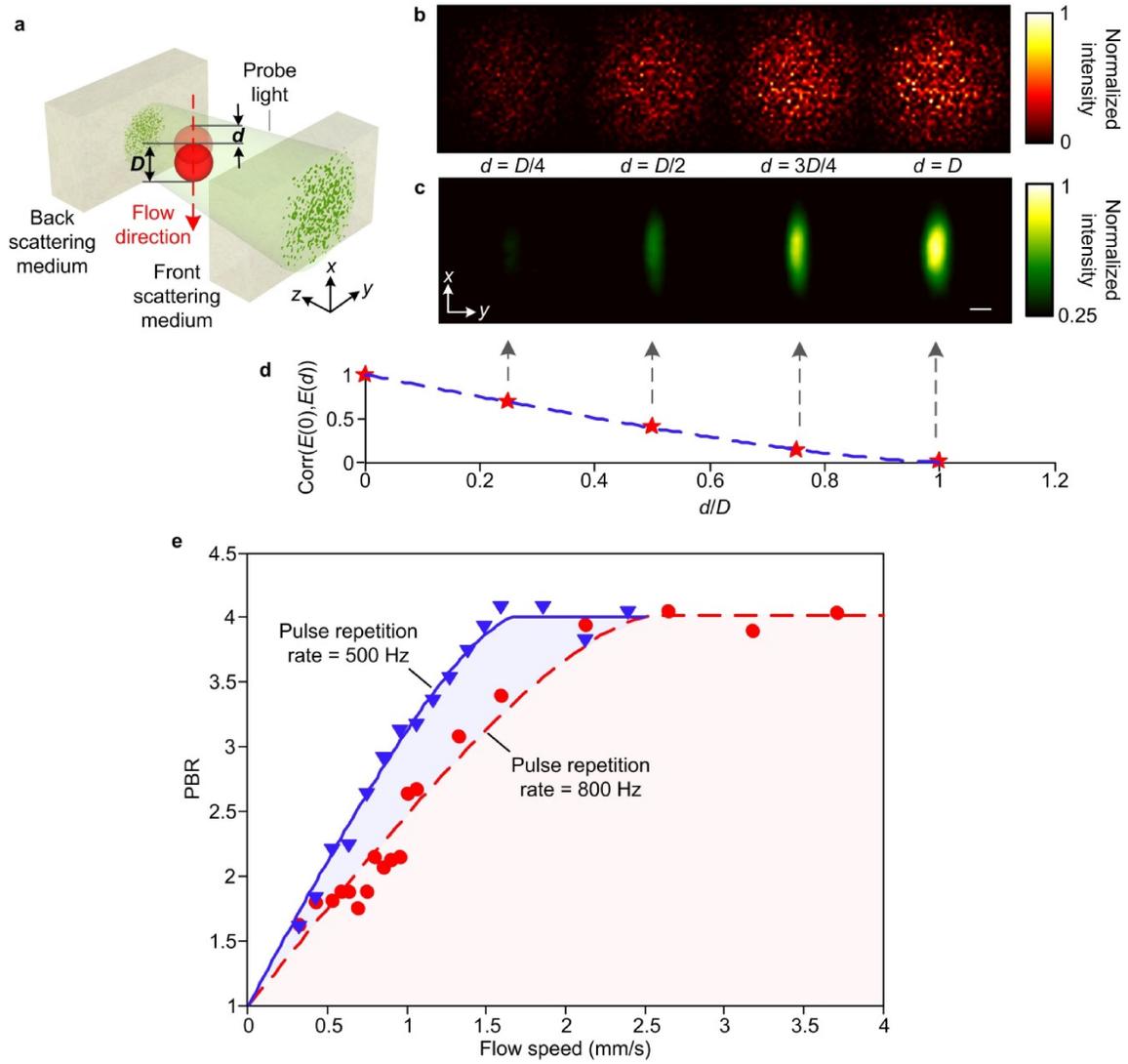